\documentclass [11pt,letterpaper]{JHEP3}
\usepackage{amsmath}
\usepackage{epsfig}
\usepackage{cite}       
\usepackage{bm}         
\usepackage{verbatim}   
\usepackage{psfrag}     
\advance\parskip 1.0pt plus 1.0pt minus 2.0pt   
\def\href#1#2{#2}	

\def\coeff#1#2{{\textstyle {\frac {#1}{#2}}}}
\def\half{\coeff 12}
\def\C{{\cal C}}
\def\P{{\cal P}}
\def\T{{\cal T}}
\def\N{{\cal N}}
\def\None{\N\,{=}\,1}
\def\Nfour{\N\,{=}\,4}
\def\R{{\mathbb R}}
\def\Rep{{\cal R}}
\def\tr{{\rm tr}}
\def\Re{{\rm Re}\,}
\def\x{\mathbf x}
\def\Nc{N_{\rm c}}
\def\Nc{N}
\def\Nf{n_{\rm f}}
\def\Z{{\mathbb Z}}

\def\Dslash{{\rlap{\raise 1pt \hbox{$\>/$}}D}}
\def\O{{\cal O}}
\def\Otilde{\widetilde \O}
\def\Veff{V_{\rm eff}}
\def\adj{\mathrm {adj}}
\def\Im{\mathrm {Im}}

\preprint{BUHEP-06...}
\preprint{}

\title
    {%
    \boldmath (In)validity of large $\Nc$ orientifold equivalence
    }%

\author
    {%
    Mithat \"Unsal$^1$\footnote{\email{unsal@buphy.bu.edu}}~
    and Laurence G.~Yaffe$^2$\footnote{\email{yaffe@phys.washington.edu}}
    \\${}^1$Department of Physics, Boston University, Boston, MA 02215
    \\${}^2$Department of Physics, University of Washington,
    Seattle, Washington 98195--1560
    }%

\abstract
    {%
    It has been argued that the bosonic sectors of
    supersymmetric $SU(N)$ Yang-Mills theory, and of QCD with a
    single fermion in the antisymmetric (or symmetric) tensor representation,
    are equivalent in the $\Nc\to\infty$ limit.
    If true, this correspondence can provide useful insight
    into properties of real QCD (with fundamental representation fermions),
    such as predictions [with $O(1/\Nc)$ corrections]
    for the non-perturbative vacuum energy,
    the chiral condensate, and a variety of other observables.
    Several papers asserting to have proven this
    large $\Nc$ ``orientifold equivalence'' have appeared.
    By considering theories compactified on $\mathbb R^3 \times S^1$,
    we show explicitly that this large $\Nc$ equivalence fails
    for sufficiently small radius, where our analysis is reliable,
    due to spontaneous symmetry breaking of charge conjugation symmetry
    in QCD with an antisymmetric (or symmetric) tensor representation fermion.
    This theory is also chirally symmetric for small radius,
    unlike super-Yang-Mills.
    The situation is completely analogous to
    large-$\Nc$ equivalences based on orbifold projections:
    simple symmetry realization conditions are both necessary and
    sufficient for the validity of the large $\Nc$ equivalence.
    Whether these symmetry realization conditions are satisfied
    depends on the specific non-perturbative dynamics
    of the theory under consideration.
    Unbroken charge conjugation symmetry is necessary for
    validity of the large $\Nc$ orientifold equivalence.
    Whether or not this condition is satisfied on $\mathbb R^4$
    (or $\mathbb R^3 \times S^1$ for sufficiently large radius)
    is not currently known.
    }%


\begin{document}

\section{Introduction and results}
\label{sec:intro}

Non-perturbative equivalences
between different gauge theories,
in the limit of a large rank $N$ of the gauge group,
can provide valuable insight
into the dynamics of strongly coupled theories.
Examples of such equivalences include volume independence
\cite{Eguchi:1982nm,
Bhanot:1982sh,
Narayanan:2003fc,
Narayanan:2005en,KUY4},
relations between $U(N)$, $SO(N)$ and $Sp(N)$ theories
\cite{Lovelace:1982hz},
relations between lattice theories with mixed fundamental/adjoint actions
\cite{Chen:1981ut, Caneschi:1981ik, Makeenko:1981bb},
and equivalences between theories related by orbifold projections
\cite{Bershadsky:1998cb,
Schmaltz:1998bg,
Erlich:1998gb,
Strassler:2001fs,
Kovtun:2003hr,
Kovtun:2004bz,
Kovtun:2005kh}.
In particular, equivalences relating supersymmetric and non-supersymmetric
theories which are confining and asymptotically free hold promise for
allowing one to convert knowledge about supersymmetric theories into
improved understanding of QCD or QCD-like theories.

There has been much discussion in the string theory literature
of orbifold and orientifold projections relating
maximally supersymmetric $\Nfour$ Yang-Mills theory to various
less supersymmetric or non-supersymmetric theories.
(See, for example, Refs.\cite{Kachru:1998ys,Lawrence:1998ja,Bershadsky:1998mb,
Adams:2001jb}.)
This prompted the observation that,
for a wide class of projections,
planar diagrams of ``parent'' and ``daughter'' gauge theories
are identical
in the $\Nc\to\infty$ limit \cite{Bershadsky:1998cb},
implying coinciding perturbative expansions of the two theories.
This planar diagram equivalence is essentially kinematic,
and does not depend on supersymmetry, conformal symmetry,
or any detailed dynamical properties of the theories.
Several authors
\cite{Schmaltz:1998bg, Strassler:2001fs, Erlich:1998gb}
suggested that a genuine non-perturbative large
$\Nc$ equivalence may exist between theories related by orbifold projections.
However, necessary and sufficient conditions for the
validity of such a non-perturbative equivalence were not clear.
Together with P.~Kovtun, we recently demonstrated that,
for a wide class of projections,
validity of large $\Nc$ equivalence between parent and daughter theories
depends {\em only\/}
on certain symmetry realizations conditions \cite{Kovtun:2004bz}.
Comparison of the $\Nc=\infty$ dynamics generated by suitable gauge invariant
coherent states shows that
the dynamics within the ``neutral'' sectors of the parent and daughter
theories coincides in the $\Nc\to\infty$ limit.
Here, neutral operators in the parent theory are gauge invariant
single-trace operators
which are invariant under the discrete symmetries used to define the
projection,
while neutral operators in the daughter theory are single-trace
gauge invariant operators
which are also invariant under global symmetries in the daughter which are
remnants of gauge symmetries in the parent theory.%
\footnote
    {
    Non-neutral operators (or states) are often called ``twisted''.
    }
This equivalence of $\Nc=\infty$ dynamics within respective neutral sectors
implies simple relations between connected correlation functions
(as well as expectation values) of corresponding neutral operators,
{\em provided} the ground (or thermal equilibrium) states of both
theories lie within their neutral sectors.
In other words, in order for this large $\Nc$ equivalence to be useful,
it is both necessary and sufficient that
{\em
neither parent nor daughter theory spontaneously break the
discrete symmetries which define their respective neutral sectors.}

In a series of papers, Armoni, Shifman, and Veneziano
assert that there is a large $\Nc$ equivalence
between the bosonic sectors of
$\None$ supersymmetric Yang-Mills (SYM) theory
and QCD with a single Dirac fermion in the
antisymmetric (or symmetric) tensor representation
\cite{Armoni:2003gp,Armoni:2003fb,Armoni:2004uu,Armoni:2003yv}.
(See also Refs.~\cite{Patella:2005vx, Barbon:2005zj, Sannino:2003xe}.)
This latter theory will be abbreviated as QCD(AS) or QCD(S)
in the antisymmetric or symmetric case, respectively.
The relation between these theories was termed an
``orientifold equivalence''
although, strictly speaking, neither theory is an orientifold
projection of the other.
This is an intriguing conjecture, with a variety of interesting
implications which have been explored in these works.
For example, validity of this equivalence implies that
QCD(AS/S), to leading order in large $\Nc$,
has vanishing vacuum energy,
a fermion condensate identical to that in $\None$ SYM,
and degenerate parity doublets.
The equivalence also implies identical patterns of
spontaneous symmetry breaking between the orientifold partners.

Armoni {\em et al.} 
\cite{Armoni:2003gp,Armoni:2004uu}
argue that the orientifold equivalence,
unlike large $\Nc$ equivalences based on orbifold projections,
is free of twisted (or non-neutral) sectors, and
therefore cannot fail due to unwanted symmetry breaking.
They claimed to have provided a rigorous non-perturbative
proof of this equivalence \cite{Armoni:2004ub}.
These authors also  assert that their construction may be
realized in a non-tachyonic string theory background and
argue, based on conventional string theory wisdom,
that this implies that the equivalence must be valid.

In this paper, we reexamine the relation between SYM
and QCD(AS/S) and reach strikingly different conclusions.
(Henceforth, ``SYM" will always mean $\None$ supersymmetric Yang-Mills theory.)
We first discuss how both $U(\Nc)$ SYM and $U(\Nc)$ QCD(AS)
theories may be obtained from $SO(2\Nc)$ SYM
by performing genuine projections.%
\footnote
    {
    A similar construction starting with $Sp(2\Nc)$ SYM
    yields $U(\Nc)$ SYM and QCD(S).
    For simplicity of presentation, we will focus on the QCD(AS) case.
    }
Viewing the relation between $U(\Nc)$ SYM and QCD(AS)
as a possible ``daughter-daughter'' equivalence
[in contrast to the more familiar parent-daughter orbifold equivalences]
clarifies the relation between these theories.
Most importantly, there {\em are} twisted sectors
in both SYM and QCD(AS)
corresponding to charge-conjugation odd states (or operators).
Of course, the mere presence of a twisted sector does not imply
any failure of large $\Nc$ equivalence,
what is important is whether the symmetry defining this
sector is spontaneously broken.
In other words, the validity of large $\Nc$ ``orientifold equivalence''
between QCD(AS) and SYM requires unbroken charge conjugation
symmetry in both theories.
This condition has not previously been noted.%
\footnote
    {
    In the discussion \cite{Armoni:2004ub,Armoni:2003gp}
    of Armoni {\em et al.},
    it was tacitly presumed that expectation values defined by the
    functional integral satisfy both charge conjugation invariance
    and large $N$ factorization.
    However, large $N$ factorization is only valid in equilibrium
    states which represent pure phases and satisfy cluster decomposition.
    If symmetries are spontaneously broken, then the functional integral
    corresponds to a mixed state which does not satisfy large $N$
    factorization.
    Patella \cite{Patella:2005vx} demonstrates
    coinciding strong coupling expansions in lattice regulated versions
    of these theories but,
    as discussed in Refs.~\cite{Kovtun:2003hr,Kovtun:2004bz},
    this only implies equivalence within the phase of the lattice theory
    which is continuously connected to arbitrarily strong coupling
    and large fermion mass.
    It does not prove equivalence outside of this phase.
    In particular,
    due to the generic presence of large $\Nc$ phase transitions
    separating weak and strong coupling,
    equivalence within the strong-coupling, large-mass phase,
    at $\Nc=\infty$, does not yield any information about continuum
    limits of these theories.
    }

Spontaneous breaking of charge conjugation symmetry
in large $N$ QCD(AS) is a non-trivial question.
There is no theorem (analogous to the Vafa-Witten theorem 
\cite{Vafa:1983tf}
for parity symmetry)
demonstrating that charge conjugation symmetry
cannot be broken spontaneously.
When the theory is formulated on $\mathbb R^4$,
the symmetry realization
depends on the long distance, strongly coupled dynamics of the theory.%
\footnote
    {
    Some results from lattice simulations of one-flavor QCD which,
    for $\Nc\,{=}\,3$, is the same as QCD(AS),
    have recently been reported \cite{DeGrand:2006uy}.
    This study used relatively small lattices and antiperiodic boundary
    conditions for the fermions,
    and did not attempt to test for spontaneous breaking of charge
    conjugation symmetry.
    Our results, presented below, show that the combination of
    small volume and antiperiodic boundary conditions suppresses
    charge conjugation symmetry breaking.
    It would be desirable to perform further lattice simulations of this theory
    and explicitly test the charge conjugation symmetry realization.
    }
One may, however, compactify on $\mathbb R^3 \times S^1$
with the circumference of the $S^1$ sufficiently small so that
the theory is weakly coupled on this scale.
In this regime, a reliable perturbative analysis is feasible.
Choosing periodic boundary conditions for the fermions,
we compute the one-loop effective potential for the Wilson line
which wraps the $S^1$.
The imaginary part of the Wilson line is an order parameter for
charge conjugation symmetry.
By analyzing the minima of the effective potential,
we find that QCD(AS)
spontaneously breaks charge conjugation invariance,
but does not break the discrete chiral symmetry of the theory.%
\footnote
    {
    Our treatment closely parallels the analysis of D.~Tong
    \cite{Tong:2002vp}, who examined a $\Z_2$ orbifold
    projection of $U(2\Nc)$ SYM, yielding a non-supersymmetric
    $U(\Nc) \times U(\Nc)$ theory, and found that
    on $\mathbb R^3 \times S^1$ with sufficiently small radius,
    the non-supersymmetric daughter theory breaks the $\Z_2$ symmetry
    exchanging gauge group factors.
    }

In contrast, supersymmetric Yang-Mills theory
is known to spontaneously break its non-anomalous
$\Z_{2\Nc}$ discrete chiral symmetry down to $\Z_2$
(corresponding to fermion number modulo two),
leading to $\Nc$ degenerate vacuum states \cite{Witten:1982df}.
Every vacuum state is invariant under a (suitably defined)
charge conjugation symmetry.
Compactifying on $\mathbb R^3 \times S^1$, for any size $S^1$,
does not change this symmetry realization.

Therefore, at least on $\mathbb R^3 \times S^1$ with small radius,
QCD(AS) and SYM have fundamentally different symmetry realizations.
We also find directly that the leading $O(\Nc^2)$ vacuum energy
is non-vanishing for QCD(AS), unlike SYM.
Therefore, when compactified with sufficiently small radius,
the previously asserted large $\Nc$ equivalence between SYM
and QCD(AS) is {\em false}.
Completely analogous results hold for compactification on
a three-torus, $T^3 \times \R$.
This disproves any possibility of a general proof of
large $\Nc$ orientifold equivalence
(independent of spacetime dimension, spatial volume, etc.).

Imposing antiperiodic boundary conditions for fermions is the same
as considering these theories at non-zero temperature.
For sufficiently high temperature,
we find that QCD(AS/S) has unbroken charge conjugation and chiral symmetry,
but spontaneously breaks its $\Z_2$ center symmetry,
indicating the deconfined nature of the high temperature phase.
Supersymmetric Yang-Mills theory, at high temperatures, also has
unbroken charge conjugation and chiral symmetry, while spontaneously
breaking its $U(1)$ center symmetry.
Therefore, large $\Nc$ equivalence of QCD(AS/S) and $\None$ SYM,
within their respective neutral sectors,
does hold at sufficiently high temperatures.

Hence, the phase diagram of QCD(AS), as a function of
temperature and spatial periods, must have one (or more)
phase transitions separating a high temperature charge conjugation
invariant deconfined phase,
in which large $\Nc$ equivalence to SYM is valid,
from a confining small volume low temperature phase
with broken charge conjugation symmetry
and no large $\Nc$ equivalence to SYM.
Whether QCD(AS) possesses a distinct low temperature large volume
confining phase in which charge conjugation is unbroken,
and large $\Nc$ equivalence to SYM is valid, is not currently known.
This is precisely the same as the status of the large $\Nc$ 
equivalence based on the $\Z_2$ orbifold projection of SYM which
yields a $U(\Nc)\times U(\Nc)$ gauge theory
with a bifundamental fermion.%
\footnote
    {
    This large $\Nc$ orbifold equivalence is applicable at sufficiently
    high temperatures, but is known to fail on $\mathbb R^3 \times S^1$
    (with periodic boundary conditions)
    for sufficiently small radius,
    due to $\Z_2$ symmetry breaking in the daughter theory.
    No evidence demonstrating $\Z_2$ symmetry breaking 
    in this specific daughter theory, in large volume,
    is known \cite{Kovtun:2005kh},
    but neither is there any proof of the absence of such symmetry breaking.
    }

To summarize, the status of the large $\Nc$ ``orientifold equivalence''
discussed in Refs.~%
\cite{Armoni:2003gp,Armoni:2003fb,Armoni:2004uu,Armoni:2003yv,
Patella:2005vx, Barbon:2005zj, Sannino:2003xe}
is no better (or worse)
than that of similar large $\Nc$ orbifold equivalences.
For all large $\Nc$ equivalences based on orbifold
or orientifold projections, appropriate symmetry realization conditions
are an unavoidable, and non-trivial, necessary condition for a useful
equivalence.

\section{Orientifold projections and daughter-daughter equivalence}
\label{sec:oneparent}

Let $\C$ denote charge conjugation.
As noted in the Introduction, QCD(AS) and $\None$ SYM 
are not directly related to each other by an orientifold projection.%
\footnote
    {
    We define orientifold projections as in Polchinski (vol 1),
    pgs.~190--192 \cite{Polchinski:1998rq}.
    Orientifold projections are $\Z_2$ projections based on a discrete
    symmetry involving charge conjugation.
    Starting with a complex group $U(N)$,
    a projection by $\C$ yields the real group $SO(N)$,
    while a projection by $\C$ combined with an antisymmetric $\Z_2$
    gauge transformations yields the pseudo-real group $Sp(N)$.
    The neutral sector in the parent $U(N)$ theory consists
    of gauge-invariant ${\cal C}$-even operators, while
    ${\cal C}$-odd operators form the twisted sector.
    There also exist ``reverse projections,''
    involving projections by suitable $\Z_2$ gauge transformations,
    which take a real $SO(2N)$ [or pseudoreal $Sp(2N)$]
    gauge group to a complex $U(N)$ group \cite{KUY4}.
    In this case, the charge conjugation symmetry of the daughter theory
    is a remnant of a $\Z_2$ transformation which is part of the gauge group
    in the parent theory.
    In this case, it is the (complex) daughter theory which has
    a twisted sector consisting of $\C$-odd operators,
    while the neutral sector contains only $\C$-even operators.
    }
However, both these theories may be obtained by applying (different)
$\Z_2$ projections to a common parent theory, namely
$SO(2\Nc)$ $\None$ super-Yang-Mills.
The field content of the theory consists of
a gauge boson $A_{\mu}$ and a Majorana gluino $\lambda$.
The parent theory has a discrete $\Z_{4\Nc-4}$ chiral symmetry,%
\footnote
    {
    See, for example, Ref.~\cite{Terning:2003th}, pg. 72.
    }
which is the non-anomalous remnant of $U(1)_R$ symmetry.
This $\Z_{4\Nc-4}$ symmetry is spontaneously broken down to the $\Z_2$
of $(-1)^F$
(corresponding to fermion number modulo two)
by the formation of a gluino condensate.
Therefore, the only unbroken internal global symmetry of
the parent theory is $(-1)^F$,
which defines the grading into fermionic and bosonic states.

Two nontrivial $\Z_2$ projections may be applied to
the parent $SO(2\Nc)$ theory which lead to $U(\Nc)$ gauge theories.
Let $J \equiv i\sigma_2 \otimes 1_N$ denote the
symplectic form which is real, antisymmetric and an element of $SO(2\Nc)$.
If one projects by $J$
({\em i.e.}, imposes the constraints
$A_\mu = J \, A_\mu \, J^T$ and
$\lambda = J \, \lambda \, J^T$ on the gauge field and gluino),
then the result is a $U(\Nc)$ gauge theory with an adjoint representation
Majorana fermion, which is precisely $U(N)$ $\None$ super-Yang-Mills theory.
Alternatively, if one projects by $J$ times $(-1)^F$
(corresponding to the constraints
$A_\mu = J \, A_\mu \, J^T$ and $\lambda = -J \, \lambda \, J^T$),
then the result is a $U(\Nc)$ gauge theory with a Dirac fermion
in the antisymmetric tensor representation,
or in other words, $U(N)$ QCD(AS).%
\footnote
    {
    Similarly, if one starts with $Sp(2\Nc)$ $\None$ SYM
    and performs similar $\Z_2$ projections,
    one may obtain either $U(\Nc)$ $\None$ SYM or $U(\Nc)$ QCD(S).
    }

\begin{FIGURE}[t]
{
  \parbox[c]{\textwidth}
  {
  \begin{center}
  \includegraphics[width=2.6in]{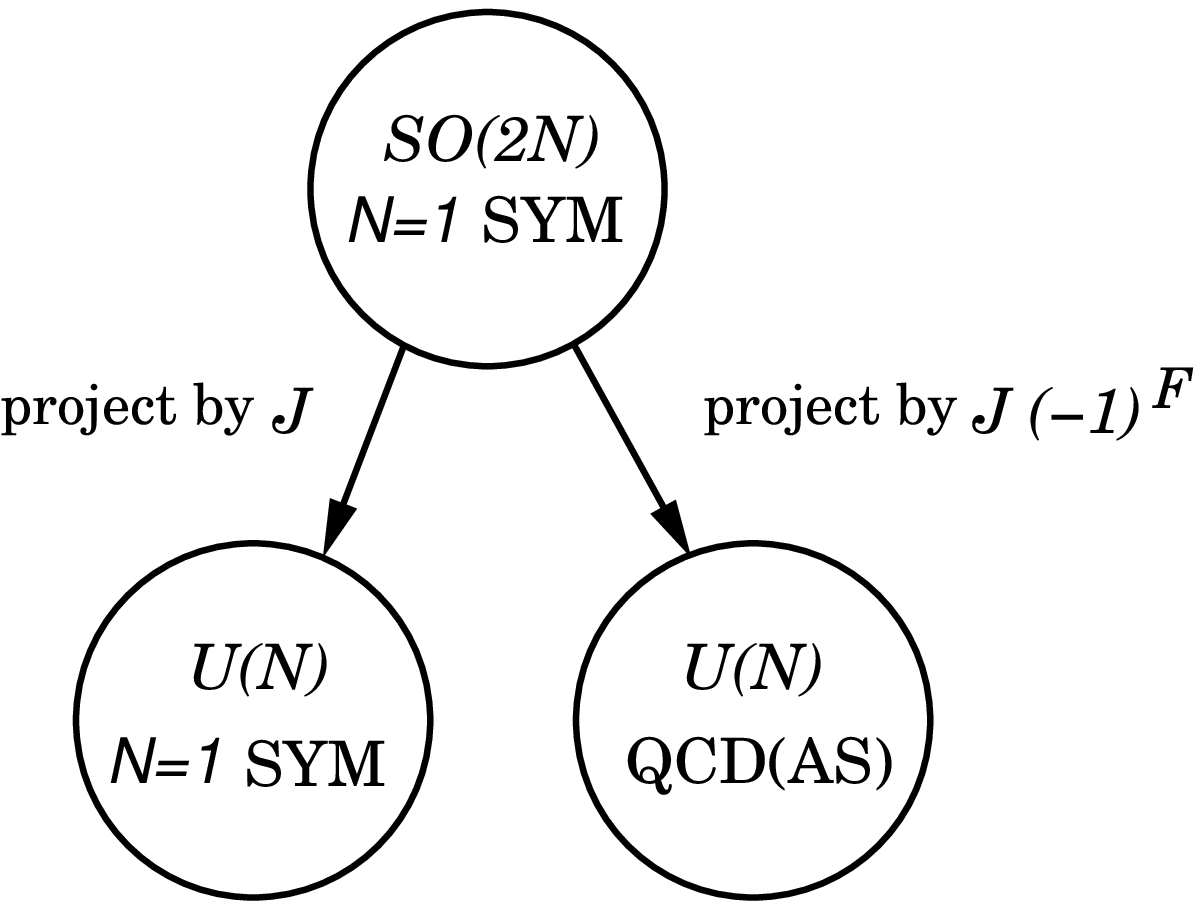}
  \caption
    {%
    $\Z_2$ projections of $SO(2\Nc)$ $\None$ super-Yang-Mills theory,
    yielding either $U(\Nc)$ SYM, or $U(\Nc)$ QCD(AS).
    Both daughter theories have a charge conjugation symmetry
    which is a remnant of a $\Z_2$ gauge transformation in the parent theory;
    hence both daughter theories have twisted sectors corresponding to
    charge conjugation odd states.
    }
  \end{center}
  }
\label{fig:projections}
}
\end{FIGURE}

We want to identify, correctly, the twisted and neutral sectors
associated with each of these projections.
Consider first the projection by $J$
yielding $U(\Nc)$ SYM
(the left branch of the figure).
This projection only involves an element of the gauge group,
and gauge symmetries never break spontaneously.
Therefore, the neutral sector in the parent $SO(2\Nc)$ theory consists
of all (gauge invariant, single-trace) operators, both fermionic and bosonic.
All such operators have images in the daughter $U(\Nc)$ theory.
The daughter theory has a complex gauge group,
and a charge conjugation symmetry $\C$.
This charge conjugation symmetry is the image of the
gauge transformation $K \equiv \sigma_3 \otimes 1_N$ in the
parent $SO(2\Nc)$ theory.
This $\Z_2$ symmetry of the daughter
is the analogue of the discrete symmetry cyclically permuting equivalent
gauge group factors in quiver theories arising from orbifold constructions.
The neutral sector in the daughter theory consists of all
(single-trace, gauge invariant) 
$\C$-even operators, both bosonic and fermionic,
while the twisted sector is the complementary set of all $\C$-odd operators.

Now consider the projection by $J (-1)^F$ yielding $U(\Nc)$ QCD(AS)
(the right branch of the figure).
The inclusion of $(-1)^F$ in the projection
modifies the twisted and neutral sectors associated with this projection.
In the parent theory, gauge invariant, single-trace operators which are
invariant under $J (-1)^F$ must be bosonic
({\em e.g.}, $\tr\,\lambda\lambda$),
while the twisted sector is composed of single-trace fermionic operators
({\em e.g.}, $\tr\,\lambda\lambda \lambda_{\alpha}$).
As in the previous case,
the daughter theory has a charge conjugation symmetry $\C$
which can be used to grade operators and is the image of
the gauge transformation by $K$ in the parent theory.
In the daughter theory,
because the fermions are in the two index antisymmetric representation
there are no gauge invariant fermionic operators
({\em i.e.}, physical operators built from an odd number of
fermion fields).
Consequently, the neutral sector in the daughter now consists of
$\C$-even bosonic operators,
while the twisted sector is composed of $\C$-odd bosonic operators.
Note that the neutral sectors, in both parent and daughter,
for the projection by $J (-1)^F$ are just the bosonic subsets
of the neutral sectors for the first projection by $J$.

In Ref.~\cite{Kovtun:2004bz} we discussed how,
by exploiting the infinite dimensional group structure which
underlies gauge invariant coherent states,
one may compare the $\Nc=\infty$ dynamics of parent
and daughter theories.
In that paper, we focused on a particular class of orbifold projections
which generate quiver gauge theories.
However, the comparison may trivially be adapted to the
projections by $J$, or $J (-1)^F$, of $SO(2\Nc)$ SYM discussed above.
Since there are no significant changes, we refer readers to
Ref.~\cite{Kovtun:2004bz} for details.
One finds exactly the same conclusion for these projections:
the $\Nc=\infty$ dynamics of parent and daughter theories
coincides {\em within their respective neutral sectors}.
In other words, the large $\Nc$ dynamics (generated by gauge invariant,
single trace operators) in $SO(2\Nc)$ SYM coincides with the
dynamics within the $\C$-even sector of $U(\Nc)$ SYM.
And the large $\Nc$ dynamics in the bosonic sector of $SO(2\Nc)$ SYM
coincides with the large $\Nc$ dynamics within the $\C$-even bosonic
sector of $U(\Nc)$ QCD(AS).

This large $\Nc$ equivalence between a common parent theory
and two different daughter theories automatically implies
an equivalence between the two daughter theories:
the large $\Nc$ dynamics of $U(\Nc)$ SYM and QCD(AS),
within their $\C$-even bosonic sectors, coincides.
This is the ``orientifold equivalence'' of Armoni {\em et al.},
and it is naturally
viewed as an example of large $\Nc$ ``daughter-daughter'' equivalence.
As the approach of Ref.~\cite{Kovtun:2004bz} makes clear,
this large $\Nc$ equivalence of SYM and QCD(AS),
within the $\C$-even bosonic sector,
is essentially kinematic.
But the utility of this equivalence depends crucially on whether
ground (or equilibrium) states of both theories lie within this sector.
If they do, then the large $\Nc$ equivalence not only implies
relations between expectation values of corresponding ($\C$-even, bosonic)
operators,
it also implies that the leading large $\Nc$ behavior
of connected correlators of such operators coincide.%
\footnote
    {
    Explicitly,
    $
	\lim_{\Nc\to\infty}
	(\Nc^2)^{M-1}
	\langle \O_1 \cdots \O_M \rangle_{\rm conn}^{\rm SYM}
	=
	\lim_{\Nc\to\infty}
	(\Nc^2)^{M-1}
	\langle \Otilde_1 \cdots \Otilde_M \rangle_{\rm conn}^{\rm QCD(AS)}
	\,,
    $
    where $\{ \O_i \}$ are $\C$-even single-trace bosonic operators
    in $U(\Nc)$ SYM and $\{ \Otilde_i \}$ are the corresponding
    ($\C$-even, bosonic, single-trace) operators produced by the
    mappings connecting both theories to $SO(2\Nc)$ SYM.
    }
But if charge conjugation symmetry is broken in either theory,
so that the ground (or equilibrium) state is not $\C$-invariant,
then the large $\Nc$ equivalence within the neutral sectors
generates no useful information about vacuum (or equilibrium)
expectation values or correlation functions.%

\section{Symmetry realizations on \boldmath $\R^3 \times S^1$ at small radius}
\label{sec:Orientifold}

In strongly coupled gauge theories there is, in general, no easy way
to tell whether a global symmetry is spontaneously broken or not.
For both QCD(AS) and $\None$ SYM,
it is possible to perform lattice simulations,
but so far only limited results on relatively small lattices
are available \cite{DeGrand:2006uy,Fleming:2000fa}.
Larger simulations with a variety of lattice sizes will be needed
to firmly establish the symmetry realization of QCD(AS) on $\R^4$.

However,
by considering this theory on $\R^3 \times S^1$ with
the circumference $L$ (or radius $L/2\pi$)
of the $S^1$ circle much smaller than the
inverse of the non-perturbative confinement scale $\Lambda$,
we can take advantage of the asymptotic freedom of the theory.
In this regime, $L \ll \Lambda^{-1}$,
the gauge coupling on the scale $L^{-1}$
is small, and perturbative methods are reliable.

\subsection*{One-loop effective potential}

We consider $\None$ SYM and QCD(AS) on $\R^3 \times S^1$
with periodic boundary conditions for the fermions on the $S^1$.
For SYM, this choice of boundary conditions preserves supersymmetry.
Let $\Omega(\x)$ denote the group valued Wilson line wrapping the $S^1$
({\em i.e.}, the path-ordered exponential of the line integral of the
gauge field around the periodic direction),
sitting at point $\x$ in $\R^3$.
Minima of the classical gauge field action correspond to vanishing
gauge field strength and constant but arbitrary values of $\Omega$.
At one-loop order, quantum fluctuations generate a non-trivial
effective potential for $\Omega$ which we will compute.%
\footnote
    {
    This exercise is a simple adaptation of the corresponding
    calculation for hot QCD in appendix D of Ref.~\cite{Gross:1980br}.
    }

We may work in a gauge where the gauge field is constant and
the Wilson line is diagonal,
\begin{equation}
    \Omega(\x) = \Omega \equiv
    {\rm diag} \, ( e^{i v_1}, \cdots,  e^{iv_{\Nc}} ) \,.
\label{Eq:background}
\end{equation}
The angles $\{ v_i \}$ are periodic variables defined modulo $2 \pi$.
The resulting one-loop effective potential, for either theory,
may be expressed as
\begin{equation}
    \Veff(\Omega) \equiv -\frac{1}{L \, \mathcal V} \, \ln Z[\Omega]
    =
    -\frac{1}{L \, \mathcal V} \,
    \ln \left[
    \frac{ {\rm det}^{\alpha}(-D_\Rep^2) } { {\rm det}({-D_\adj^2}) }
    \right] ,
\label{eq:Veff}
\end{equation}
where $D_\Rep^2$ denotes the covariant Laplacian,
in the background of a constant gauge field,
for representation $\Rep$,
and $\mathcal V$ is the volume of $\R^3$.
The functional determinant in the numerator of the logarithm
comes from integrating out fermions in representation $\Rep$, while
the adjoint representation determinant in the denominator is the
combined result of fluctuations in the gauge boson and ghost fields.
The exponent $\alpha$ equals 1 for a Majorana fermion
(relevant for $\None$ SYM),
while $\alpha = 2$ for a Dirac fermion [relevant for QCD(AS)].

   From the form (\ref{eq:Veff}), it is immediate that the one-loop
effective potential vanishes identically for $\None$ SYM
(whose Majorana fermion is in the adjoint representation),
\begin{equation}
    \Veff^{\rm SYM}(\Omega) = 0 \,.
\end{equation}
As is well known,
due to supersymmetry
this remains true to all orders in perturbation theory.
(Non-perturbative effects generate a non-vanishing potential.
This is reviewed below.)

For QCD(AS), one must compute the functional determinants in the
adjoint and antisymmetric tensor representations.
Details are given in the Appendix.
One finds,%
\footnote
    {
    For QCD(S), the only difference is an additional contribution
    of
    $
	-2  \sum_{i=1}^{\Nc} \, [2v_i]^2 ( 2\pi - [2v_i] )^2
	+ \coeff {16}{15} \pi^4 \Nc
    $,
    inside the braces of (\ref{Eq:potential}),
    from the diagonal components of the symmetric representation fermions.
    This only affects the subleading $O(\Nc)$ term to the effective potential
    and has no effect on the leading large $\Nc$ dynamics.
    Hence, the following discussion of QCD(AS) applies equally well to QCD(S).
    }
\begin{eqnarray}
    \Veff^{\rm QCD(AS)}(\Omega) &=& \frac{1}{24\pi^2L^4} \>
    \Big\{
      \sum_{i,j=1}^{\Nc}
	  [v_i {-} v_j ]^2 \left( 2\pi - [v_i {-} v_j] \right)^2
	- \coeff {8}{15} \, \pi^4 \Nc
\nonumber\\ && \qquad{}
      - 2 \sum_{i<j=1}^{\Nc}
	  [v_i {+} v_j ]^2 \left( 2\pi - [v_i {+} v_j] \right)^2
    \Big\} \,,
\label{Eq:potential}
\end{eqnarray}
up to higher order corrections suppressed by $g^2$.
Here $[x] \equiv x \bmod 2\pi$ indicates quantities defined to lie
within the interval $[0, 2\pi)$.
The first sum is the contribution of gauge bosons (and ghosts),
and is the same as the result one finds in hot gauge theories.
The second sum is the contribution of fermions and,
due to the use of periodic instead of antiperiodic boundary conditions,
differs from the thermal compactification result.
The $O(\Nc)$ constant term reflects the imperfect cancellation
of the zero point energies between the $\Nc^2$ components of the gauge
field, and the $\Nc(\Nc{-}1)$ components of the fermions.
This is a subleading $O(1/\Nc)$ correction relative to the leading
$O(\Nc^2)$ contributions of the sums.

From the result (\ref{Eq:potential}), one sees that
the gluon contribution is positive definite
and is minimized when the eigenvalues of $\Omega$ coincide,
so that $v_i = v_j \pmod {2\pi}$.
In other words, this term generates an effective
attraction between eigenvalues.
On the other hand,
the fermionic contribution is negative definite.
Global minima of the effective potential are necessarily located
on the subspace where all eigenvalues coincide,
$v_i =v$,
since within this subspace one can simultaneously minimize every
term in the first sum, and maximize every term in the second sum.
Within this  subspace, the potential (\ref{Eq:potential}) equals
\begin{equation}
    \Veff^{\rm QCD(AS)}(e^{iv})
    =
    - \frac{\Nc(\Nc{-}1) }{24\pi^2 L^4} \>
    [2v]^2 \left( 2\pi - [2v] \right)^2
    -
    \frac{\pi^2 \Nc }{45 \, L^4} \,.
\label{Eq:potential1}
\end{equation}
This function, which is plotted in Fig.~\ref{fig:Potential},
has two global minima at
\begin{equation}
    v= \pi/2  \quad  {\rm and}\quad v= 3\pi/2 \,,
\end{equation}
or in other words, when $\Omega = \pm i$ (times the unit matrix).
Charge conjugation acts on the Wilson line $\Omega$ as complex conjugation,
or equivalently sends $v_i \to 2\pi - v_i$.
The potential (\ref{Eq:potential1}) is charge conjugation symmetric,
but its two minima are exchanged by the action of $\C$.

\begin{FIGURE}[t]
{
  \parbox[c]{\textwidth}
  {
  \begin{center}
  \includegraphics[width=3.0in]{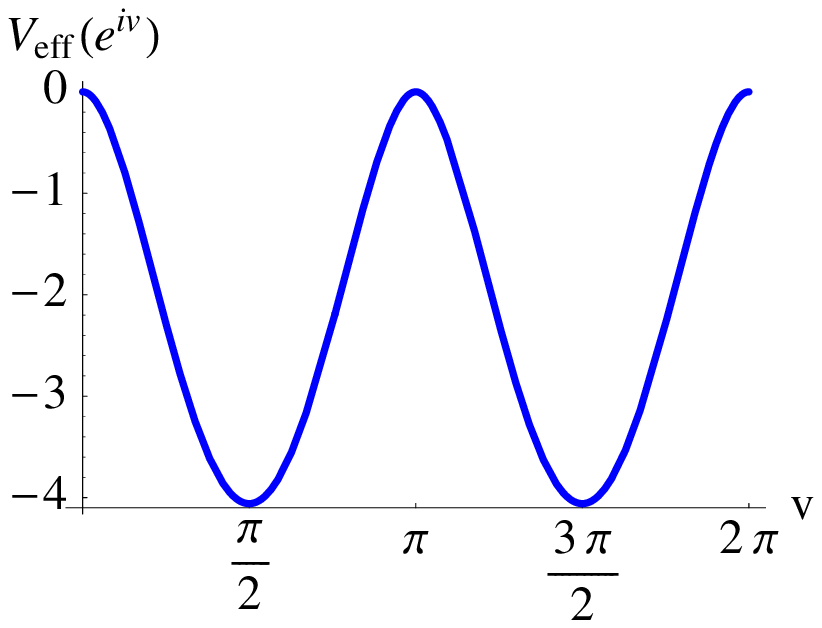}
  \caption
    {%
    The one loop effective potential in QCD(AS/S),
    minus the irrelevant $O(N)$ additive constant and
    divided by $N(N{-}1)/(\pi^2 L^4)$,
    as a function of the phase $v$ of the Wilson line
    (when all eigenvalues coincide).
    There are two minima at $v = \pi/2$ and $3\pi/2$.
    These two minima are exchanged by charge conjugation $\C$,
    indicating that $\C$ breaks spontaneously in QCD(AS) on
    $\R^3 \times S^1$ when the circumference $L$ is small
    compared to $\Lambda^{-1}$.
    For ${\cal N}=1$ SYM, the one loop potential identically vanishes.
    }
  \end{center}
  }
\label{fig:Potential}
}
\end{FIGURE}

If $L \ll \Lambda^{-1}$, so that $g^2(1/L) \ll 1$, then higher
order corrections to the effective potential are small perturbations
which cannot change the conclusion that the effective potential
has two distinct gauge-inequivalent minima, related by charge conjugation.
This shows that QCD(AS) on $\R^3 \times S^1$, for sufficiently small radius,
spontaneously breaks charge conjugation symmetry.
The imaginary part of the trace of the Wilson line is an order parameter
for this symmetry.
Its expectation value,
\begin{equation}
    \Bigl\langle \Im \, \frac{\tr \> \Omega}\Nc \Bigr\rangle
    = \pm i + O(g^2) \,,
\end{equation}
is nonzero and demonstrates spontaneous  breaking of $\C$.
The vacuum energy density equals the value of the effective potential
at its minima and, to lowest order, is%
\footnote
    {
    $\Omega = \pm i$
    is the correct vacuum configuration for
    $U(\Nc)$ QCD(AS) theory for any value of $\Nc$.
    Careful readers will notice that this configuration 
    is only an element of $SU(\Nc)$ when $\Nc$ is a multiple of 4.
    For $SU(\Nc)$ theories with $\Nc \bmod 4 \ne 0$,
    vacuum configurations are those elements of the center of $SU(\Nc)$
    which lie closest to $\pm i$.
    For large $\Nc$, the resulting vacuum energy
    difference between $SU(\Nc)$ and $U(\Nc)$ is of order one,
    and hence subleading relative to the $O(\Nc^2)$ result (\ref{eq:Evac}).
    Change conjugation symmetry in $SU(\Nc)$ QCD(AS)
    (on $\R^3 \times S^1$ with small radius)
    is spontaneously broken for all values of $\Nc > 2$.
    \label{fn:SU(N)}
    }
\begin{equation}
    {\cal E} = -  \frac{\pi^2 }{24 \, L^4} \> \Nc(\Nc{-} \coeff 7{15}) \,.
\label{eq:Evac}
\end{equation}

In contrast, the vacuum structure of the $\None$ SYM theory, on
$\mathbb R^3 \times S^1$, is essentially  dictated by supersymmetry.
At the perturbative level, to all orders, $\Veff(\Omega) =0$,
implying a compact moduli space of $T^N/S_N$
on which the eigenvalues of $\Omega$ roam freely.
[$T^N$ is the maximal torus
and $S_N$ the Weyl group of $U(N)$.]
Nonperturbatively, however, the moduli space is lifted by
instanton and monopole contributions \cite{Affleck:1982as}.
A nonperturbative superpotential is generated,
and it provides a repulsive interaction between eigenvalues.
Consequently, as shown in a nice treatment by Davies {\em et al.}
\cite{Davies:1999uw, Davies:2000nw},
in the vacuum state, the eigenvalues
of the Wilson line are distributed equidistantly around the unit circle.
Both the vacuum energy, and $\langle \tr \, \Omega \rangle$ vanish identically.
Charge conjugation symmetry [as well as the $U(1)$ center symmetry]
is unbroken in $\None$ SYM for all values of the $S^1$ circumference.

\subsection*{Chiral condensates}

It is also instructive to compare the chiral properties of QCD(AS/S) and
$\None$ SYM.
The case of $\None$ SYM is well known \cite{Witten:1982df}.
The theory has a $\Z_{2\Nc}$ chiral symmetry which is the
non-anomalous discrete remnant of the anomalous $U(1)_R$ symmetry.
This $\Z_{2\Nc}$ chiral symmetry is spontaneously broken down to the $\Z_2$
of $(-1)^F$ by the formation of a fermion bilinear condensate.
As a result, the theory has $\Nc$ isolated, supersymmetric vacua.
This symmetry realization is independent of the size of the $S^{1}$ circle,
and remains valid in the decompactified $L\to\infty$ limit.
Each of the $\Nc$ vacua are invariant under a suitably redefined
charge conjugation symmetry (equal to $\C$ times one of the elements
of the $\Z_{2\Nc}$ chiral symmetry).

\begin{FIGURE}[ht]
{
  \parbox[c]{\textwidth}
  {
  \begin{center}
  \includegraphics[width=3.6in]{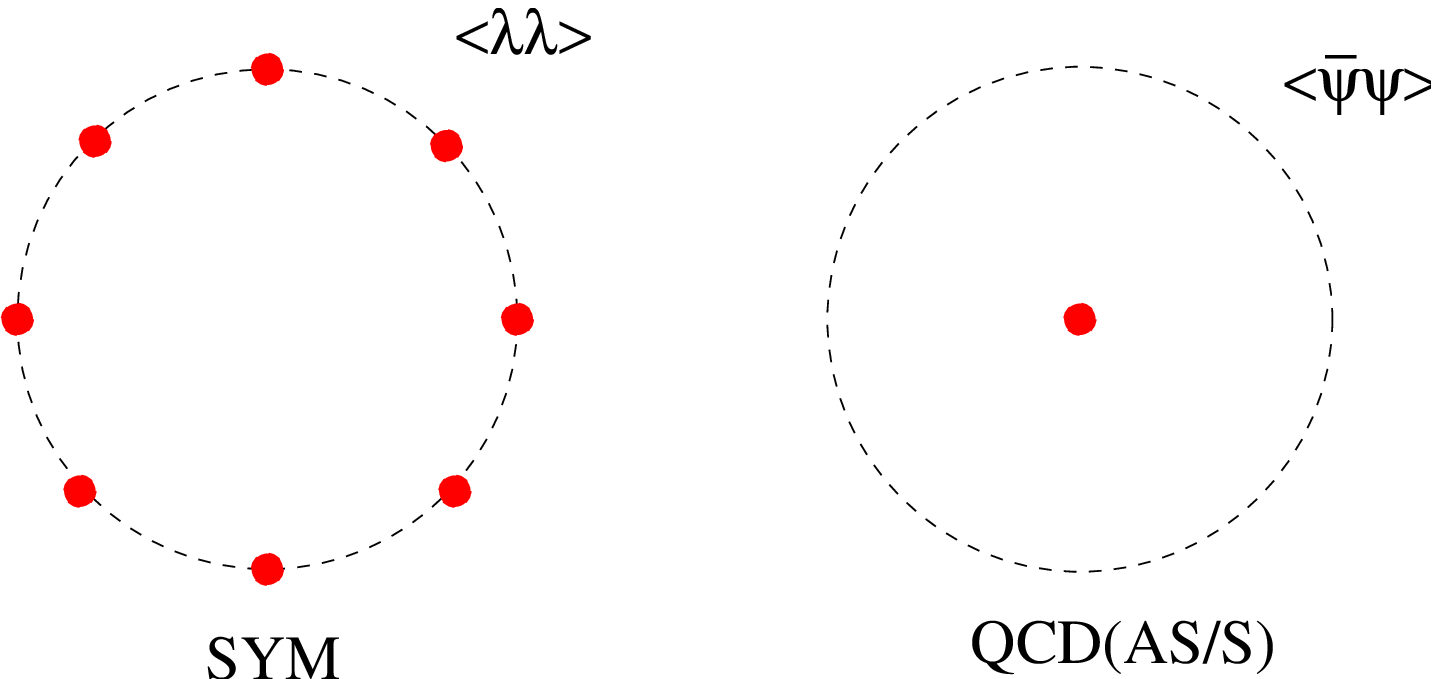}
  \caption
    {%
    Chiral condensates of $\None$ SYM (left) and QCD(AS/S) (right),
    for $L \ll \Lambda^{-1}$.
    The $N=1$ SYM theory has $\Nc$ chirally asymmetric vacuum states in which
    the fermion condensate $\langle \lambda\lambda \rangle$
    has a non-zero magnitude and a phase equal to an integer multiple
    of $2\pi/\Nc$.
    In contrast, QCD(AS/S) has a chirally
    symmetric phase with vanishing chiral condensate.
    }
  \end{center}
  }
\label{fig:chiral}
}
\end{FIGURE}

QCD(AS) has an analogous $\Z_{2\Nc-4}$ 
non-anomalous discrete chiral symmetry
[and QCD(S) has a $\Z_{2\Nc+4}$ chiral symmetry].
(See, for example, Ref.~\cite{Terning:2003th}.)
On $\R^3 \times S^1$ with small circumference, $L \ll \Lambda^{-1}$,
this chiral symmetry is unbroken, unlike the case for $\None$ SYM.
To see this, note that the value $\Omega = \pm i$ for the  Wilson line,
corresponding to one of the minima of the effective potential discussed above,
is equivalent to a constant background gauge field $A_4 = \pm \pi/(2L)$
(with $x_4$ labeling the compactified direction).
This has the effect of shifting the allowed frequencies in the
mode decomposition of antisymmetric
representation fermions from the usual integer multiples of
$2\pi/ L$,
appropriate for periodic boundary conditions and no gauge field,
to integers plus a half,
$\omega_n = 2\pi (n + \half)/L$, $n = 0, \pm 1, \pm 2, \cdots$.
(It is a shift by $1/2$, not $1/4$, because the
fermions are in a two-index representation.)
Hence, all allowed frequencies are non-zero and bounded below,
in magnitude, by $\pi/L$.
When analyzing dynamics at distances large compared to $L$,
the fermion field may be viewed as a Kaluza-Klein tower of three-dimensional
fields, with effective masses equal to $|\omega_n|$.
All modes act like very heavy fields
and may be integrated out perturbatively,
with no formation of any non-perturbative fermion condensate
and consequently no breaking of chiral symmetry.%
\footnote
    {
    Another way to understand this is to note that
    fermion propagators fall exponentially like $e^{-\pi|\x|/L}$
    at large distances, $|\x| \gg L$.
    Therefore, fermionic correlations are exponentially small
    at distances where the gauge field
    fluctuations become strongly coupled.
    Chiral symmetry breaking, if it were to occur,
    would be signaled by the failure of cluster decomposition
    in the chirally-invariant functional integral measure,
    $
	\lim_{|\x|\to\infty}
	\langle {\cal O}(\x)^\dagger {\cal O}(0) \rangle
	\ne 0
    $
    for ${\cal O}(\x)$ a chiral symmetry order parameter
    such as $\bar\psi(\x) \psi(\x)$.
    Such a correlator must involve two (or more) fermion propagators
    connecting the two operators.
    If fermion propagators are exponentially small at large
    distance, then this violation of cluster decomposition
    cannot possibly occur.
    For a rigorous proof along these lines,
    in the context of high temperature lattice gauge theory,
    see Refs.~\cite{TomboulisYaffe1,TomboulisYaffe2}.
    }

Therefore, when the compactification size $L$ is small compared to
$\Lambda^{-1}$, QCD(AS/S)
spontaneously breaks charge conjugation symmetry but does not break
discrete chiral symmetry,
while $\None$ SYM does exactly the opposite, breaking chiral symmetry
but not charge conjugation.
This difference in chiral symmetry realizations is illustrated
in Fig.~\ref{fig:chiral}.

\subsection*{Symmetry realizations and order parameters}

In addition to charge conjugation
(whose realization determines whether
there is a useful large $\Nc$ equivalence between QCD(AS/S) and SYM)
and discrete chiral symmetry,
it is also instructive to consider the realization of other global symmetries.
Compactifying one direction preserves the parity invariance which
is a symmetry of both $\None$ SYM and QCD(AS/S).
The compactification also creates a new global symmetry known
as center symmetry.%
\footnote
    {
    Let $\hat x$ denote the compactified direction (with period $L$).
    Center symmetry is the invariance of a gauge theory
    under gauge transformations $g(x)$ which are only periodic
    up to an element of the center of the gauge group,
    $g(x+L \, \hat x) = \omega \, g(x)$,
    with $\omega$ some element of the gauge group
    (other than the identity) which commutes with all group elements.
    Although most easily described as invariance under aperiodic
    gauge transformations, 
    center symmetry should be regarded as a {\em global\/} symmetry
    of the theory.
    Physical states need not be invariant under non-trivial
    center symmetry transformations;
    Gauss' law only requires that physical states be invariant
    under periodic gauge transformations (which are continuously
    connected to the identity).
    The global center symmetry is really the quotient of the
    full gauge group (including aperiodic gauge transformations)
    divided by the subgroup of periodic transformations.
    Since all physical states are invariant under periodic gauge
    transformations, one may identify non-trivial center symmetry
    transformations which only differ by a periodic gauge transformation.
    }
For $U(\Nc)$ SYM, the center symmetry is a continuous $U(1)$ invariance.
For $U(\Nc)$ QCD(AS/S), the presence of fermions with $\Nc$-ality two
reduces the center symmetry to $\Z_2$, corresponding to invariance
under gauge transformations which are antiperiodic in the compactified
direction.
For the following discussion,
let $\cal Z$ denote an antiperiodic center symmetry transformation.

Both charge conjugation $\C$ and parity $\P$
(or $x \to -x$ reflection) map the Wilson line to its conjugate,
$\Omega \to \Omega^\dagger$,
while the center symmetry transformation $\cal Z$ of QCD(AS/S)
negates the Wilson line,
$\Omega \to -\Omega$.
Therefore,
the imaginary part of the trace of the Wilson line,
$\Im (\tr \, \Omega)$,
is odd under each of the three symmetry transformations
$\C$, $\P$, and $\cal Z$,
but is even under the product of any two of these transformations.
Consequently, when the expectation value of $\Im(\tr\,\Omega)$
is non-zero, this corresponds to a symmetry breaking pattern
in which the $(\Z_2)^3$ symmetry group generated by $\C$, $\P$,
and $\cal Z$ is spontaneously broken down to the $(\Z_2)^2$ subgroup
consisting of $\C\P$, $\cal CZ$, and $\cal PZ$ (plus the identity),
resulting in two degenerate vacua, as seen in the above analysis
of the effective potential.

The only operators which can function as order parameters for this particular
symmetry breaking pattern must,
like $\Im (\tr \, \Omega)$,
be odd under each of
$\C$, $\P$, and $\cal Z$.
Being odd under only one or two of these transformations is not sufficient.%
\footnote
    {
    An order parameter must transform non-trivially under at least some
    of the spontaneously broken symmetries
    while being invariant under the unbroken symmetry group.
    If an operator $\cal O$ is invariant under any one of the
    three transformations $\C$, $\P$, and $\cal Z$,
    and is also invariant under the unbroken
    $\cal CP$, $\cal CZ$ and $\cal PZ$ transformations,
    then it is necessarily invariant under each of
    $\cal C$, $\cal P$, and $\cal Z$.
    }
Consequently, generic $\C$-odd operators,
such as the imaginary part of topologically trivial Wilson loops,
cannot reveal this symmetry breaking pattern.
Only non-local $\C$, $\P$, and $\cal Z$ odd operators,
such as $\Im (\tr\, \Omega^k)$, for $k$ odd,
will have expectation values which reveal
the symmetry breaking in QCD(AS/S) at small radius.

There is one more symmetry worth discussing, namely time-reversal $\T$.
This transformation leaves invariant the trace of the spacelike Wilson,%
\footnote
    {
    For reference, if the gauge field is regarded as an
    anti-Hermitian matrix and the metric signature is
    $({-}{+}{+}{+})$, then the actions of $\C$, $\P$
    and $\T$ transformations on the gauge field are given by
    $
	A_\mu(t,{\bf x}) \mathop{\to}\limits^\C +A_\mu(t,{\bf x})^*
    $,
    $
	A_\mu(t,{\bf x}) \mathop{\to}\limits^\P -A^\mu(t,-{\bf x})
    $,
    and
    $
	A_\mu(t,{\bf x}) \mathop{\to}\limits^\T +A^\mu(-t,{\bf x})
    $.
    }
$
    {\T} \, (\tr\, \Omega) \, {\T}^{-1}
    =
    \tr \, \Omega \,.
$
Hence one might think that an imaginary expectation value for
$\tr \,\Omega$
would have no implications for the realization of time reversal symmetry.
This is incorrect, however,
because time-reversal is an anti-unitary transformation.
If one decomposes $\tr \,\Omega$ into real and imaginary parts,
then
$
    \Re (\tr \,\Omega)
    = \frac 1{2} \left[ \tr \,\Omega + \tr \,\Omega^\dagger \right]
$
is time-reversal even, while
the Hermitian operator
$
    \Im (\tr \,\Omega) =
    \frac 1{2i}\left[ \tr \,\Omega - \tr \,\Omega^\dagger \right]
$
is necessarily time-reversal odd.
Unbroken time reversal symmetry implies that any Hermitian,
time-reversal odd operator must have vanishing expectation value.
Consequently, in addition to signaling spontaneous symmetry
breaking of charge conjugation, parity, and center symmetry,
the non-zero expectation value of $\Im (\tr \,\Omega)$
in QCD(AS/S), at small radius, also implies spontaneous
breaking of time reversal (and hence also CPT) symmetry.
But, as noted above, the only observables which are sensitive
to this symmetry breaking are operators involving
topologically non-trivial Wilson loops
with odd winding numbers around the compactified direction.

\subsection*{High temperature}

Changing the fermion boundary conditions on the $S^1$ 
from periodic to antiperiodic
allows the theory on $\R^3 \times S^1$ to be reinterpreted as
a thermal field theory on $\R^3$ space with
inverse temperature $\beta = L$.
The Wilson line $\Omega(\x)$ wrapping the $S^1$ is now a thermal
Polyakov loop,
and serves as an order parameter for the center symmetry of the theory.%
\footnote
    {
    When the compactified direction is regarded as Euclidean time
    (with antiperiodic boundary conditions for fermions),
    then the center symmetry realization determines
    whether the theory is in a confining or deconfined phase.
    Unbroken center symmetry means a confining phase, while
    spontaneous breaking of center symmetry indicates a deconfined phase
    \cite{centersym1,centersym2}.
    }
Once again, if the temperature $T = L^{-1}$ is large compared to the
confinement scale $\Lambda$, then a perturbative calculation of the
effective potential for the Wilson line is reliable.
For QCD(AS/S),
changing the boundary conditions for the fermions is exactly equivalent
to replacing $\Omega$ by $i \Omega$ in the result (\ref{Eq:potential})
for the effective potential,
or equivalently shifting the eigenvalues $v_i \to v_i + \pi/2$.
(Switching the sign in the boundary conditions corresponds to
a shift of $\pi/2$, not $\pi$, because the fermions are in a
two index representation.)
Hence, the one-loop potential for QCD(AS) becomes
\begin{eqnarray}
    \Veff^{\rm QCD(AS)}(\Omega) &=& \frac{T^4}{24\pi^2} \,
    \Big\{
      \sum_{i,j=1}^{\Nc}
	  [v_i {-} v_j ]^2 \left( [v_i {-} v_j]  - 2 \pi \right)^2
	  - \coeff 8{15} \, \pi^4 \Nc
\nonumber\\ && \quad {}
      - 2 \sum_{i<j=1}^{\Nc}
	  [v_i {+} v_j {+}\pi]^2 \left( [v_i {+} v_j{+}\pi] - 2 \pi \right)^2
    \Big\} \,.
\label{Eq:potential2}
\end{eqnarray}
This is now minimized when all $v_i = 0$, or all $v_i = \pi$,
or in other words, when $\Omega = \pm 1$.
This demonstrates spontaneous breaking of the $\Z_2$ center symmetry,
and indicates that the high temperature phase of the theory is
a deconfined phase.
The value of this potential at its minimum gives the leading order
free energy density (or minus the pressure),
\begin{equation}
    \mathcal F^{\rm QCD(AS)}
    =
    -\frac {\pi^2 }{24} \> T^4\,\Nc (\Nc - \coeff 7{15} )
    \,,
\label{eq:FQCDAS}
\end{equation}
which equals the expected Stefan-Boltzmann result of
$-\frac{\pi^2}{45} \> T^4 \left[ \Nc^2 +  \frac 78 \, \Nc(\Nc{-}1) \right]$.

The non-vanishing Polyakov loop,
$\Omega = \pm 1$,
and doubly degenerate minima
signal spontaneous breaking of the $\Z_2$ center symmetry of QCD(AS).
But because $\langle \tr \, \Omega \rangle$ is now real,
there is no longer any spontaneous
breaking of charge conjugation symmetry (or parity or time-reversal)
in this high temperature
deconfined phase.%
\footnote
    {
    Although we have only performed a perturbative analysis,
    these conclusions are clearly valid non-perturba\-tively.
    As for ordinary QCD,
    the equilibrium long-distance non-perturbative dynamics of
    high temperature QCD(AS/S) is described by a three-dimensional
    effective theory which is just 3$d$ Yang-Mills.
    This theory has a unique ground state and no symmetry breaking.
    }
Chiral symmetry is also unbroken in this phase.

Since charge conjugation symmetry is unbroken in this hot plasma phase,
the large-$\Nc$ equivalence to $\None$ SYM, within the neutral sector,
is guaranteed to be applicable to thermal expectation values
and connected correlators.%
\footnote
    {%
    In the case of the nonsupersymmetric ${\Z_2}$ orbifold projection
    of $\None$ SYM \cite{Kovtun:2005kh},
    precisely the same considerations 
    reveal that the $\Z_2$ symmetry permuting gauge group factors
    is unbroken in the high temperature, deconfined phase of the theory,.
    Therefore, large-$\Nc$ equivalence to $\None$ SYM is 
    applicable within this phase of the theory.
    }
However, it may be instructive to see this explicitly
in a simple example, such as the large $\Nc$ free energy.

Thermal (antiperiodic) boundary conditions for the fermions
break supersymmetry and cause SYM to develop a non-vanishing
effective potential for the Polyakov loop.
Evaluating the required functional determinant for adjoint
representation fermions with antiperiodic boundary conditions,
as described in the Appendix, leads to
\begin{equation}
    \Veff^{\rm SYM}(\Omega) = \frac{T^4}{24\pi^2} \,
    \Big\{
      \sum_{i,j=1}^{\Nc}
	  [v_i {-} v_j ]^2 \left( [v_i {-} v_j]  - 2 \pi \right)^2
      - \sum_{i,j=1}^{\Nc}
	  [v_i {-} v_j {+}\pi]^2 \left( [v_i {-} v_j{+}\pi] - 2 \pi \right)^2
    \Big\} \,.
\label{Eq:potential3}
\end{equation}
This is minimized when all eigenvalues coincide, or when
$\Omega = e^{iv} \, \rlap 1 \, 1$ for any phase $v$.
This demonstrates spontaneous breaking of the $U(1)$ center symmetry
in hot SYM.
The resulting leading order free energy density is
\begin{equation}
    \mathcal F^{\rm SYM}
    =
    -\frac {\pi^2}{24} \> T^4 \, \Nc^2 \,,
\label{eq:FSYM}
\end{equation}
which is the expected Stefan-Boltzmann result for this theory.
As required by large $\Nc$ equivalence, the leading $O(\Nc^2)$
piece of the QCD(AS) free energy (\ref{eq:FQCDAS}) coincides
with the SYM result (\ref{eq:FSYM}).
At this lowest order, the comparison is essentially trivial,
and just reflects that both theories have, to order $\Nc^2$,
the same number of bosonic and fermionic degrees of freedom.
But the large-$\Nc$ equivalence guarantees that the $O(\Nc^2)$
parts of the free energy, in the high temperature phase,
will coincide exactly.

Spontaneous breaking of the center symmetry generates
an uncountable number of extremal equilibrium states
(labeled by the phase $v$) in $U(\Nc)$ SYM,
since the center symmetry is a continuous $U(1)$,
but only two extremal states in QCD(AS/S),
where the center symmetry is a discrete $\Z_2$.
Only for the two states of hot SYM with
$\langle \tr \, \Omega/N \rangle = \pm 1$
do the Polyakov loop expectation values coincide 
with QCD(AS/S).
This is completely consistent with the large $\Nc$ equivalence,
as these are the only extremal equilibrium states of SYM
which lie within the neutral (charge conjugation invariant) sector
to which the large $\Nc$ equivalence applies.

\subsection*{Compactification on \boldmath $T^3 \times {\mathbb R}$}

Instead of compactifying a single direction one may, of course,
choose to compactify two or more directions.
A particular case, which has been previously considered by
Barbon and Hoyos \cite{Barbon:2005zj},
is compactification on $T^3 \times {\mathbb R}$,
with a symmetric three torus and periodic boundary conditions
for the fermions in all directions.
For this finite volume spatial compactification,
all dynamics is weakly coupled if the physical size of the three torus
is much smaller than confinement scale $\Lambda^{-1}$.
For QCD(AS) (as well as the $\Z_2$ orbifold projection of SYM),
Ref.\cite{Barbon:2005zj} analyzed the resulting one-loop effective potential
for Wilson lines.
This corresponds to a Born-Oppenheimer approximation in the
quantum mechanics of these finite volume theories.

The discussion in Ref.~\cite{Barbon:2005zj} was focused on trying
to identify properties of these non-supersymmetric theories which might
somehow be related, at large $\Nc$, to the supersymmetric index of SYM.
It did not directly address the much more basic issue of the symmetry
realizations of these theories in the $\Nc\to\infty$ limit.

Examination of the effective potential found in Ref.~\cite{Barbon:2005zj}
for QCD(AS) shows that it has eight-fold degenerate global minima
corresponding to Wilson lines along each of the three directions of the torus
equaling $\pm i$.
In the $\Nc \to \infty$ limit, this is properly interpreted as
indicating spontaneous breaking of both the $(\Z_2)^3$ center symmetry
on the torus, and charge conjugation symmetry.%
\footnote
    {
    For finite (spatial) volume and finite values of $\Nc$,
    the theory must have a unique ground state and no spontaneous
    symmetry breaking (just like a generic quantum theory with
    a finite number of degrees of freedom).
    But the $\Nc \rightarrow \infty $ limit is a thermodynamic limit,
    just like the infinite volume limit of a typical statistical system,
    and spontaneous symmetry breaking in the $\Nc=\infty$ limit
    is perfectly possible.
    The tunneling amplitudes between different minima
    of the effective potential decrease exponentially with increasing $\Nc$,
    so the lifetime of a state localized near a single minimum
    diverges as $\Nc\to\infty$.

    More formally,
    one may test for spontaneous symmetry breaking in either of two ways.
    One may add a symmetry breaking perturbation of strength $\epsilon$
    to the theory and test whether the $\epsilon \to 0$ limit
    of the expectation value of an order parameter,
    after sending $\Nc \to \infty$,
    depends on the direction of approach of $\epsilon$ to zero.
    Alternatively, one may leave the theory unchanged,
    so that the ground state is completely symmetric,
    and instead test whether expectation values of products of order
    parameters, in the $\Nc\to\infty$ limit, satisfy large $\Nc$ factorization.
    This corresponds to testing cluster decomposition in the
    usual large volume limit, and is an equally valid indicator
    of symmetry breaking.
    Failure of large $\Nc$ factorization (or cluster decomposition)
    implies that, in the thermodynamic limit, the symmetric ground
    state is indistinguishable from a statistical mixture of extremal
    pure states which do satisfy factorization (or cluster decomposition).
    }
In other words, the status of large $\Nc$ orientifold (or orbifold)
equivalence is exactly the same on
$T^3 \times \R$ with small volume and
$\R^3 \times S^1$ with small radius.
In either case,
symmetry breaking in the non-supersymmetric daughter theory
(with periodic boundary conditions)
prevents any useful large $\Nc$ equivalence to SYM.

\section{Phase diagrams}

Examination of the one loop effective potential has taught us that
charge conjugation symmetry is spontaneously broken in QCD(AS/S)
(at zero temperature) when one spatial dimension is compactified
with size $L\ll\Lambda^{-1}$.
However, this perturbative analysis can not tell us anything
about symmetry realizations when $L\gg \Lambda^{-1}$.
No proof demonstrating the absence
(or presence) of spontaneous breaking of charge conjugation
in the strongly coupled large radius regime is known.
For sufficiently large radius (or in the decompactified theory on $\R^4$),
QCD(AS) [or QCD(S)] is expected to develop a chiral condensate,
$\langle \bar\psi\psi \rangle \ne 0$,
and thereby spontaneously break its discrete chiral symmetry
down to the $\Z_2$ of $(-1)^F$.
This is not unequivocally established,
but is consistent with the effects of instanton
induced interactions,
and is supported by the results of the simulations reported
in Ref.~\cite{DeGrand:2006uy}.
Assuming this is the case, then QCD(AS/S) on $\R^3 \times S^1$
must have a chiral symmetry breaking phase transition at some
critical size $L_\chi$
(which would necessarily be of order $\Lambda^{-1}$).

If the chirally-asymmetric large radius phase has unbroken
charge conjugation symmetry,
then within this phase the large $\Nc$ equivalence connecting
QCD(AS/S) and $\None$ SYM is useful and, as discussed in
Refs.~\cite{Armoni:2003gp,Armoni:2003fb,Armoni:2004uu,Armoni:2003yv},
generates interesting quantitative predictions
such as equality of chiral condensates.
The simplest scenario is that there is a single phase transition
at a critical size $L_c$ where both chiral symmetry and charge conjugation
realizations change.
But it is also possible that there are two distinct transitions,
at radii $L_\chi$ and $L_\C$,
with chiral symmetry breaking for $L > L_\chi$
and broken charge conjugation symmetry for $L < L_\C$.
If so, then the large-$\Nc$ equivalence to SYM implies that
$L_\chi$ must be less than $L_\C$ (for large $\Nc$),
since otherwise the vanishing of the chiral condensate
in the interval  $L_\C < L  < L_\chi$
would contradict the large-$\Nc$ equivalence to SYM.
If there are two distinct transitions,
then the intermediate phase for $L_\chi < L < L_\C$ would have both chiral
symmetry and charge conjugation spontaneously broken,
and no useful large-$\Nc$ equivalence to SYM.

It is also logically possible that that charge conjugation
symmetry remains broken for all radii,
in which case QCD(AS) [or QCD(S)] would have a single
phase transition and a large radius phase in which both
discrete chiral symmetry and charge conjugation symmetry
are spontaneously broken.
In this case, the large $\Nc$ equivalence to $\None$ SYM,
which is valid only in the neutral sectors of these theories,
would not be useful for predicting low energy properties
since the ground state of QCD(AS) would always lie outside this sector.

Examining these theories as a function of temperature and the
compactified spatial radius reveals a richer phase structure.
(By this we mean considering these theories on $\R^2 \times (S^1)^2$,
with one $S^1$ having circumference $L$ and periodic boundary
conditions for the fermions, and the other $S^1$ having circumference
$\beta = 1/T$ and antiperiodic boundary conditions.)
As discussed in the previous section,
at sufficiently high temperatures both SYM and QCD(AS/S) must have
a deconfined plasma phase in which chiral and charge conjugation
symmetries are unbroken.
(If $\beta \ll L$, then the compactification of a spatial direction
is not significant in the analysis of high temperature thermodynamics.
Hence, the discussion of high temperature in the last section applies
equally well to this regime of two compactified directions.)

\begin{FIGURE}[ht]
{
  \parbox[c]{\textwidth}
  {
  \vspace*{-30pt}
  \begin{center}
  \psfrag{inf}{$\infty$}
  \raisebox{3pt}{\includegraphics[width=3.0in]{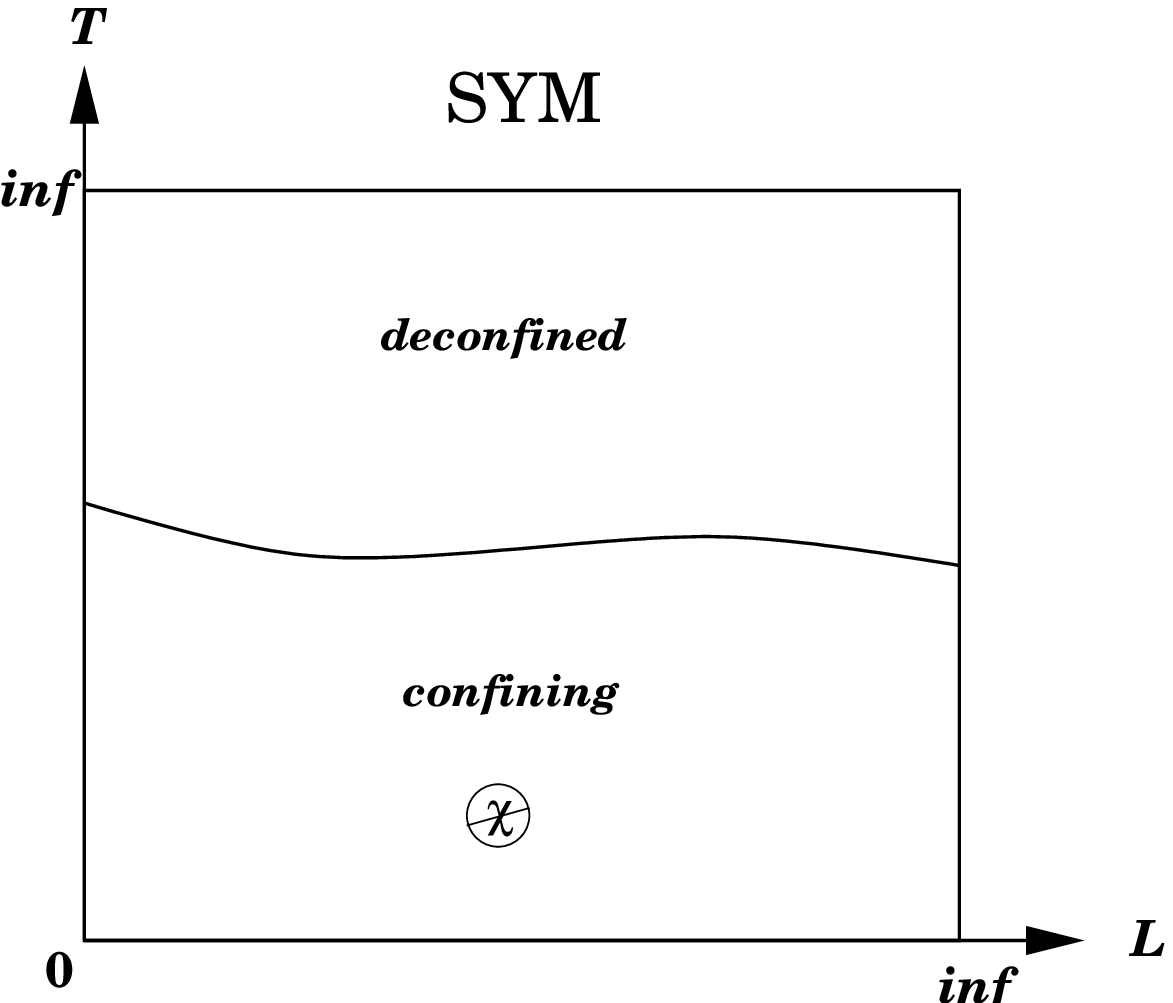}}
  \includegraphics[width=3.0in]{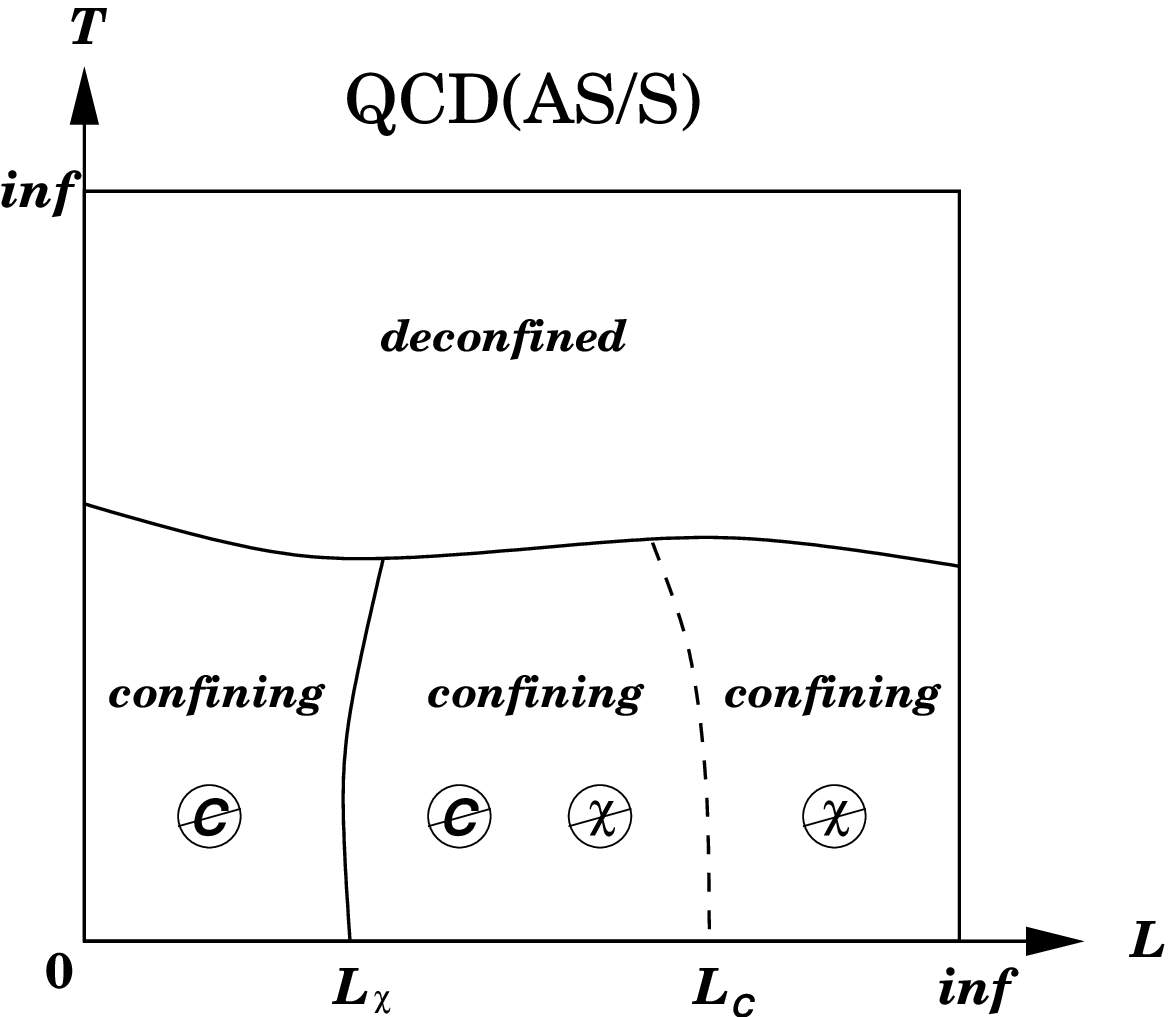}
  \vspace*{-15pt}
  \caption
    {%
    Schematic phase diagram of $U(\Nc)$ $\None$ SYM (left)
    and QCD(AS/S) (right)
    as a function of temperature $T$ and compactification size $L$
    of one spatial direction.
    $\None$ super-Yang-Mills has a confining low temperature phase
    in which the discrete chiral symmetry is spontaneously broken,
    and a deconfined high temperature phase with unbroken chiral
    symmetry.  Charge conjugation is unbroken in both phases.
    QCD(AS/S) has multiple confining low temperature phases.
    For sufficiently small radius, $L < L_\C$, it spontaneously
    breaks charge conjugation symmetry.
    For sufficiently large radius, $L > L_\chi$, it is believed
    to break spontaneously its discrete chiral symmetry.
    As discussed in the text, 
    the intermediate phase, with both charge conjugation and chiral
    symmetry breaking, may not exist, or it may extend all the way
    to $L = \infty$.
    The deconfined high temperature phase has unbroken chiral and
    charge conjugation symmetry.
    Large $\Nc$ equivalence to SYM applies only within the phases
    of QCD(AS/S) with unbroken charge conjugation symmetry.
    These sketches depict the simplest scenario in which only
    a single phase transition separates high and low temperature phases;
    more complicated scenarios with distinct deconfinement,
    chiral restoration, and [for QCD(AS/S)] charge conjugation
    restoration transitions are also possible.
    }
  \end{center}
  }
\label{fig:phase}
}
\end{FIGURE}

In $\None$ SYM, for any value of the spatial size $L$,
as one raises the temperature there must be at least one phase
transition separating the confining, chirally asymmetric low
temperature phase from the chirally symmetric high temperature
plasma phase.
[The $U(1)$ center symmetry is spontaneously broken in the plasma
phase, and unbroken in the confining phase.]
Whether there is a single phase transition, at which both
confinement and chiral symmetry realizations change,
or two phase transitions separating a distinct intermediate
phase is not clear.
We are unaware of any evidence indicating which alternative applies.

In QCD(AS/S), all the possible zero temperature phases discussed above
are confining phases.
Starting in any of these phases, as the temperature is increased
there must again be one or more phase transitions separating
the low temperature phase from a high temperature plasma phase.
(The $\Z_2$ center symmetry of the theory is spontaneously broken
in the plasma phase, and unbroken in the confining phase.)
For some values of $L$, there could be transitions at three distinct
temperatures corresponding to separate chiral symmetry restoration,
charge conjugation restoration, and deconfinement transitions.

Fig.~\ref{fig:phase} depicts the simplest possibility,
in which a single transition separates the low temperature phases 
from the high temperature plasma phase.
But as just noted, much more complicated possibilities might occur.
Large-$\Nc$ equivalence between QCD(AS/S) and $\None$ SYM
(for equilibrium correlators of neutral operators)
is valid in any phase of QCD(AS/S) which
does not spontaneously break charge conjugation symmetry ---
and only in such phases.
Hence, the equivalence is valid within the deconfined plasma phase.
And it also applies within the large radius low temperature phase with
(only) broken chiral symmetry --- if this phase exists.%
\footnote
    {
    The discussion of this section is equally applicable to the
    $\Z_2$ orbifold projection of $\None$ SYM yielding a
    non-supersymmetric
    $U(\Nc)\times U(\Nc)$ gauge theory with a bifundamental fermion.
    Charge conjugation symmetry in that case is replaced by the
    $\Z_2$ symmetry interchanging the two gauge group factors.
    With this substitution, the sketch of the
    phase diagram of QCD(AS/S) in Fig.~\ref{fig:phase}
    is equally appropriate for the $\Z_2$ orbifold case.
    }

It should be emphasized that the preceding discussion,
and Fig.~\ref{fig:phase},
are specifically addressing the case of $U(\Nc)$ gauge theories
and, for QCD(AS), tacitly assume that $\Nc > 3$.
If $\Nc = 3$, then QCD(AS) has no chiral symmetry
other than the $\Z_2$ of $(-1)^F$,
and hence no chiral symmetry breaking.
And (as noted in footnote \ref{fn:SU(N)})
spontaneous breaking of charge conjugation (and $\P$ and $\T$) symmetry
at small radius only occurs for $\Nc > 2$.%
\footnote
{
Replacing the $U(\Nc)$ gauge group by $SU(\Nc)$
eliminates the $U(1)$ photon from the theory.
In the large $\Nc$ limit, this change is irrelevant.
But for finite $\Nc$ there are some important differences.
For SYM theory, the $U(1)$ gauge field is decoupled from all
other degrees of freedom, for any $\Nc$, and removing it
has no effect of the dynamics or the phase diagram.
    [%
    Switching from $U(\Nc)$ to $SU(\Nc)$ in SYM does 
    change the center symmetry from $U(1)$ to $\Z_{\Nc}$.
    In both cases the center symmetry is spontaneously broken
    in the high temperature phase of the theory.
    This is true even when one spatial dimension is compactified,
    so that the long distance physics is effectively two dimensional.
    Hot $U(\Nc)$ SYM with its free $U(1)$ gauge field
    illustrates the fact that one can formally have spontaneous
    symmetry breaking of a continuous symmetry in two dimensions
    if the Goldstone bosons are decoupled.%
    ]
But for $SU(\Nc)$ QCD(AS/S),
only if $\Nc$ is even does the theory have a
$\Z_2$ center symmetry,
whose realization provides a sharp distinction between
confined and deconfined phases.
There is no center symmetry if $\Nc$ is odd,
reflecting the fact that, in this case,
(anti)symmetric representation fermions
can screen test charges in any representation.
Assume, for the sake of discussion, that spontaneous breaking
of charge conjugation symmetry does not survive to arbitrarily
large radius, so that $L_\C$ in Fig.~\ref{fig:phase} is finite.
If one raises the temperature starting in
the confining large volume phase of $SU(\Nc)$ QCD(AS/S) then,
for $\Nc$ even, there must be a sharp deconfinement transition
even in the presence of an non-zero fermion mass
(which eliminates the chiral symmetry and, for sufficiently large
mass, can eliminate a chiral phase transition).
(See also Ref.~\cite{Sannino:2005sk}.)
But for $\Nc$ odd, there need be no phase transition
associated with deconfinement.
The massless theory [with $\Nc > 3$ for QCD(AS)]
must still have a sharp chiral transition,
but at non-zero fermion mass this can turn into a smooth crossover.

$SU(3)$ QCD(AS) is a special case with no
center symmetry and no chiral symmetry [other than $(-1)^F$].
In this theory, for any fermion mass,
there can be a smooth crossover with no sharp transitions separating
the confining large volume low temperature phase
and the high temperature plasma.%
    {
    This is consistent with the numerical results of
    Refs.~\cite{Alexandrou:1998wv,DeGrand:2006uy}.
    }
It is not clear if this sensitivity of symmetry realizations
to the value of $\Nc$,
for $SU(\Nc)$ gauge groups, will be reflected in other
properties of QCD(AS) (such as particle spectra).
}

\section{Multiple fermion flavors} 

Instead of considering theories with just one fermion field,
one may generalize the entire previous discussion to the case
of multiple fermion flavors.
Starting with a non-supersymmetric $SO(2\Nc)$ Yang-Mills with $\Nf$
adjoint representation Majorana fermions,
orientifold projections by $J$, or $J (-1)^F$, 
yield $U(\Nc)$ gauge theories with either $\Nf$ adjoint representation
Majorana fermions,
or $\Nf$ antisymmetric tensor representation Dirac fermions, respectively.
(To preserve asymptotic freedom in these theories,
$\Nf$ must be at most five.)
Once again, there is a large-$\Nc$ equivalence between these
two daughter theories within their respective neutral sectors
(corresponding to bosonic, charge conjugation even operators).

It is completely straightforward to generalize the analysis of section
\ref{sec:Orientifold} to multiple (massless) flavors.
When compactified on $\R^3 \times S^1$ with sufficiently small radius,
and periodic boundary conditions,
one finds that the one loop effective potential for the Wilson line
in $U(\Nc)$ Yang-Mills with $\Nf$ adjoint fermions generates
a repulsive interaction between eigenvalues.
Consequently, the Wilson line eigenvalues distribute uniformly
around the unit circle and $\langle \tr \, \Omega\rangle=0$.
Charge conjugation symmetry is unbroken.
This theory has a non-anomalous $SU(\Nf) \times \Z_{2\Nc \Nf}$
chiral symmetry which is expected to break down to 
$SO(\Nf)$ by the formation of a fermion bilinear condensate,
giving rise to a vacuum manifold with $\Nc$ disjoint components,
each of which is the coset space $SU(\Nf)/SO(\Nf)$.
We expect this chiral symmetry realization to hold
for all values of the $S^1$ circumference.
The same analysis with antiperiodic boundary conditions
shows that at sufficiently high temperatures the theory has
spontaneously broken center symmetry but
unbroken chiral and charge conjugation symmetry, 
just like hot SYM.

The situation for QCD(AS) with $\Nf > 1$ flavors is essentially
the same as for $\Nf = 1$.
On $\R^3 \times S^1$ with sufficiently small radius,
charge conjugation symmetry is spontaneously broken.
Hence, the ground state does not lie in the neutral sector
and large $\Nc$ equivalence to Yang-Mills with adjoint fermions
is not applicable to vacuum expectation values.
Also, within this phase of the theory, the 
$SU(\Nf)_L \times SU(\Nf)_R \times \mathbb \Z_{(2\Nc-4) \, \Nf}$ 
chiral symmetry remains unbroken.
For sufficiently large radius,
we expect the chiral symmetry to be spontaneously broken down to
$SU(\Nf)_V$ leading to a vacuum manifold with $\Nc-2$
components each of which is the coset space
$[SU(\Nf)_L \times SU(\Nf)_R]/ SU(\Nf)_V$.
At sufficiently high temperature, it is again easy to establish that
charge conjugation and chiral symmetries are unbroken,
and large $\Nc$ equivalence to Yang-Mills with adjoint fermions,
within the neutral sector, is applicable to this phase.

The realization of charge conjugation symmetry,
at large radius (and zero temperature), is not currently known. 
Let us assume, for the sake of discussion, that charge conjugation 
symmetry is unbroken for sufficiently large radius.
Then large $\Nc$ equivalence between the multiflavor adjoint and
antisymmetric representation theories
will be applicable to their confining large radius phases.
One point to note is that the number of Goldstone bosons does {\em not\/}
match between these two theories.
For Yang-Mills with $\Nf$ adjoint fermions, there are
$\half\Nf(\Nf+1)-1$ Goldstone bosons while QCD(AS) with $\Nf$ flavors has
$\Nf^2-1$.
This is not in conflict with large $\Nc$ equivalence between these
theories, as only $\half\Nf(\Nf+1)-1$ of the Goldstone bosons of
multiflavor QCD(AS) are charge conjugation even --- and this
correctly coincides with the number of Goldstone bosons
in the multiflavor adjoint representation theory (all of which are $\C$ even).

\section{Remarks}

\subsubsection*{Orbifolds versus orientifolds, and kinematics versus dynamics}

One of the motivations in Ref.~\cite{Armoni:2003gp}
for considering the orientifold equivalence between SYM and QCD(AS/S)
was the belief that twisted sectors were absent,
coupled with a belief that the mere
existence of a twisted sector would spoil large $\Nc$ equivalence
\cite{Armoni:2003gp,Armoni:2004uu}.
(These points are also emphasized in the recent work \cite{Armoni:2005wt}.)
However, for either orbifold or orientifold equivalences,
the presence of twisted sectors is an inevitable consequence
of the projections by discrete symmetries which define the mappings
between theories.
As we have emphasized, what is significant for large $\Nc$ equivalence
is not the existence of twisted sectors, but rather the realization
of the symmetries which define the neutral and twisted sectors.
In this regard, orbifold and orientifold equivalences are on exactly
the same footing.

As illustrated by the specific example of QCD(AS/S),
it is also inevitable that the symmetry realization of a theory
depends on its specific dynamics and on the phase of the theory
under consideration.
In much of the discussion in the literature concerning ``tests''
of large $\Nc$ orbifold or orientifold equivalence,
the focus has been on kinematic aspects of the mapping between theories,
without addressing the more serious issue of the influence of dynamics
on the symmetry realization.

It is important to understand that large $\Nc$ equivalence between
theories does not mean equality.
Rather, it means that there is a well defined mapping
connecting a specific class of (neutral) observables in the two theories.
In some cases, such as orbifold projections which change
the dimension of the gauge group,
the correct mapping between theories requires appropriate
rescalings of operators and correlation functions
\cite{Kovtun:2004bz,Kovtun:2005kh}.
Misunderstanding of the correct mapping between theories,
or of the limitation of the equivalence to appropriate neutral sectors,
has led to several unjustified claims of inequivalence.

\subsubsection*{String theory realizations}

It has been argued that large $\Nc$ equivalence between $\None$ SYM
and QCD(AS/S) may naturally be understood in the context of
non-tachyonic string theory, and that this ensures the validity
of the orientifold equivalence.
(See  pages 71--80 of Ref.~\cite{Armoni:2004uu} and references therein, 
as well as  section 7 of Ref.~\cite{Armoni:2005wt}.)
This is an interesting argument, especially in light of our
demonstration that orientifold equivalence {\em can\/} fail
due to spontaneous breaking of charge conjugation symmetry.
The essence of this argument boils down to three steps:
{\em (i)} the  confining, asymptotically free gauge theories of
interest [$\None$ SYM and QCD(AS/S)] have string theory duals,
{\em (ii)} the dual string theory realizations of these theories are,
by construction, tachyon free, and
{\em (iii)} any symmetry-breaking instability in field theory would
necessarily appear as a tachyon in the string theory realization.

We believe these assertions overlook essential caveats which
undermine this argument.
There is no string theory realization of $\None$ SYM or QCD(AS/S)
for which the string theory is under control
in the decoupling limit needed to obtain precisely
the non-gravitational theories of interest.
(The recent paper \cite{Bena:2006rg}
of Bena {\em et al.} is relevant in this regard.)
Any string theory realization of an asymptotically free theory
such as SYM and QCD(AS/S) will necessarily involve a highly
curved background spacetime.
The regime of the string theory where its dynamics is under control
does not include the regime of interest.
Therefore the presence, or absence, of tachyons in a weakly-coupled
regime of the string theory, where it does not correspond to the
desired field theory,
tells one nothing about symmetry realizations in that field theory.

\subsubsection*{Summary}

Large $\Nc$ ``orientifold'' equivalence between QCD(AS/S)
and $\None$ SYM fails when the theories are compactified on small torii,
due to spontaneous breaking of charge conjugation symmetry.
The equivalence holds at sufficiently high temperature,
when charge conjugation symmetry is unbroken.
Whether charge conjugation symmetry remains spontaneously
broken in large radius compactifications (or in the decompactified limit
on $\R^4$)
is not yet clear, and depends on detailed dynamics of the theory.

This situation is exactly parallel to the case of large $\Nc$ equivalence
between $U(2\Nc)$ $\None$ SYM and its $\Z_2$ orbifold projection
yielding a $U(\Nc) \times U(\Nc)$ gauge theory with a bifundamental
fermion \cite{Kovtun:2005kh}.
This equivalence is valid at high temperature, but fails
when the theories are compactified on sufficiently small torii,
due to spontaneous breaking of the $\Z_2$ symmetry exchanging
gauge groups in the daughter theory.
Whether this symmetry is restored or remains broken in
large volume is also not currently known.

\acknowledgments
We thank  Andy Cohen,  Tom DeGrand,
Andreas Karch, Adam Martin, Takemichi Okui and 
Misha Shifman for conversations related to this work. M.\"U. thanks 
the Aspen Center for Physics where portions of this paper were completed.
This work was supported by the
U.S.\ Department of Energy under Grants DE-FG02-91ER40676
and DE-FG02-96ER-40956.

\appendix

\section{Functional determinants on \boldmath $\R^3 \times S^1$}
\label{sec:appendix}

We need to evaluate $\ln \det_\pm (-D^2_\Rep)$,
the logarithm of the functional determinant of the
covariant Laplacian
acting on $\R^3 \times S^1$
in the presence of an arbitrary constant $U(\Nc)$
gauge field pointing in the compact
direction, for various representations $\Rep$ and
with either periodic or antiperiodic boundary conditions on the $S^1$
(indicated by the subscript on the determinant).
[A single Dirac fermion gives rise to
$\det (\rlap{\,\slash} D_\Rep)$ which,
since the gauge field strength vanishes, equals $\det^2(-D^2_\Rep)$.
A Majorana fermion gives the Pfaffian (or square root of the determinant)
of $\rlap{\,\slash} D_\Rep$, which equals $\det(-D^2_\Rep)$.]

Working in a gauge in which the background gauge field
(in the fundamental representation) is diagonal,
with eigenvalues $\{ v_i/L \}$
(with $L$ the circumference of the $S^1$),
makes it easy to diagonalize the covariant Laplacian.
To compute the log of the determinant,
defined via dimensional continuation,
the key ingredient is the identity
\begin{eqnarray}
    g(v) &\equiv&
    L^3 \sum_n \int \frac {d^3 k}{(2\pi)^3} \>
    \ln \left[ k^2 + (2\pi n + v)^2 L^{-2} \right]
    =
	-\coeff {\pi^2}{45}
	+ \coeff 1{24 \pi^2} \> [v]^2 \, (2\pi - [v])^2 \,,
\end{eqnarray}
where $[v] \equiv v \bmod 2\pi$.
(For further details see, for example, appendix D of Ref.~\cite{Gross:1980br}.)

Consequently, the functional determinant of the Laplacian in the
fundamental representation is
given by
\begin{eqnarray}
    \ln {\det}_{+}(-D^2_{\rm fund}) &=&
    \frac{\mathcal V}{L^3} \sum_{i=1}^\Nc \> g(v_i) \,,
\quad
    \ln {\det}_{-}(-D^2_{\rm fund}) =
    \frac{\mathcal V}{L^3} \sum_{i=1}^\Nc \> g(v_i+\pi) \,,
\end{eqnarray}
where $\mathcal V$ is the spatial volume.
For multi-index representations, one merely has to think how the
gauge field, viewed as a diagonal $\Nc \times \Nc$ matrix, acts on the
individual components of the representation.
For the adjoint, symmetric tensor, and antisymmetric tensor representations,
the appropriate generalizations are:
\begin{eqnarray}
    \ln {\det}_{+}(-D^2_{\rm adj}) &=&
    \frac{\mathcal V}{L^3} \sum_{i,j=1}^\Nc g(v_i{-}v_j) \,,
\;\; \>
    \ln {\det}_{-}(-D^2_{\rm adj}) =
    \frac{\mathcal V}{L^3} \sum_{i,j=1}^\Nc g(v_i{-}v_j{+}\pi) \,,
\\
    \ln {\det}_{+}(-D^2_{\rm sym}) &=&
    \frac{\mathcal V}{L^3} \sum_{i \le j=1}^\Nc g(v_i{+}v_j) \,,
\;\;
    \ln {\det}_{-}(-D^2_{\rm sym}) =
    \frac{\mathcal V}{L^3} \sum_{i \le j=1}^\Nc g(v_i{+}v_j{+}\pi) \,,
\\
    \ln {\det}_{+}(-D^2_{\rm antisym}) &=&
    \frac{\mathcal V}{L^3} \!\sum_{i < j=1}^\Nc g(v_i{+}v_j) \,,
\;\;
    \ln {\det}_{-}(-D^2_{\rm antisym}) =
    \frac{\mathcal V}{L^3} \!\sum_{i < j=1}^\Nc g(v_i{+}v_j{+}\pi) .\quad\;
\end{eqnarray}

\bibliographystyle{JHEP}
\bibliography{orientifold1}

\end{document}